\documentclass[%
 aip,
 amsmath,amssymb,
 reprint,%
]{revtex4-1}

\usepackage{graphicx}
\usepackage{dcolumn}
\usepackage{bm}

\usepackage[utf8]{inputenc}
\usepackage[T1]{fontenc}
\usepackage{mathptmx}
\usepackage{etoolbox}

\makeatletter
\def\@email#1#2{%
 \endgroup
 \patchcmd{\titleblock@produce}
  {\frontmatter@RRAPformat}
  {\frontmatter@RRAPformat{\produce@RRAP{*#1\href{mailto:#2}{#2}}}\frontmatter@RRAPformat}
  {}{}
}%
\makeatother
\begin{document}

\preprint{AIP/123-QED}

\title{Model of Charge Transfer Collisions between $C_{60}$ and Slow Ions}
\author{J. Smucker}
 \affiliation{Department of Physics, University of Connecticut.}
  \email{jonathan.smucker@uconn.edu.}
\author{J. A. Montgomery Jr}
\affiliation{Department of Physics, University of Connecticut.}
\author{M. Bredice}%
 \affiliation{Department of Physics, University of Connecticut.}
 \author{M. G. Rozman}
\affiliation{Department of Physics, University of Connecticut.}
\author{R. C\^{o}t\'{e}}
\affiliation{%
Department of Physics, University of Massachusetts Boston.}
\author{H. R. Sadeghpour}
\affiliation{ITAMP, Center for Astrophysics$\,|\,$ Harvard \& Smithsonian.}
\author{D. Vrinceanu}
\affiliation{Department of Physics, Texas Southern University.}
\author{V. Kharchenko}
\affiliation{Department of Physics, University of Connecticut.}
\affiliation{ITAMP, Center for Astrophysics$\,|\,$ Harvard \& Smithsonian.}

\date{\today}

\begin{abstract}
A semi-classical model describing the charge transfer collisions of $C_{60}$ fullerene with different slow ions has been developed to explain available experimental data. This data reveals multiple Breit-Wigner like peaks in the cross sections, with subsequent peaks of reactive cross sections decreasing in magnitude. Calculations of the charge transfer probabilities and cross sections for quasi-resonant and reactive collisions have been performed using semi-empirical interaction potentials between fullerenes and ion projectiles. All computations have been carried out with realistic wave functions for $C_{60}$'s valence electrons derived from the simplified jellium model. The quality of these electron wave functions have been successfully verified by comparing theoretical calculations and experimental data on the small angle cross sections of resonant $C_{60}+ C_{60}^+$ collisions. Using the semi-empirical potentials to describe resonant scattering phenomena in $C_{60}$ collisions with ions and Landau-Zener charge transfer theory, we calculated theoretical cross sections for various $C_{60}$ charge transfer and fragmentation reactions which agree with experiments.
\end{abstract}

\maketitle

\section{Introduction}
Charge transfer in ion-atom collisions have been studied intensively for decades, mostly due to high demands in plasma physics, astrophysics, and atmospheric science \cite{AtomIonCollisionsReviewDelos,ColdAtomIonReviewMichal,CTReview}. A detailed understanding of the electron transfer processes in ion-atom collisions is at hand thanks to observations and theoretical analysis involving both quantum and classical mechanics. The most accurate results have been obtained for resonant and quasi-resonant charge transfer because only a few states are involved in the electron capture process. The majority of ion-atom systems do not support resonances between electron states and are mostly treated in expensive numerical calculations operating with extended basis functions \cite{CTTheoryRapp}. Recent examples of such multi-state calculations can be found in the theoretical analysis of charge transfer collisions of highly charged ions \cite{lyons2017charge}.

Significant simplification can be carried out for the theoretical description of slow collisions, when velocities of heavy nuclei are much smaller than typical electron velocities. In the semi-classical approximation, the motion of heavy particles is considered classically and electronic degrees of freedom are treated quantum-mechanically \cite{child1996molecular,nikitin2012theory,atomicCollisons,CollisionTheory}. In this model, transitions between different electronic states occur in specific regions of non-adiabatic behaviour, such as areas of pseudo-crossing \cite{olson1970LZCT,olson1970LZStueck,zener1932OrigLZ} or crossing (non-adiabatic transition induced by rotational interactions \cite{nikitin2012theory}) of electron energy curves.

Current development of experimental and theoretical investigations includes investigations of charge transfer collisions with more complex targets, such as molecules, nano-size clusters and condense matter materials \cite{los1990charge,ChargeTransferNanoTubesRao,ChargeTransferGraphene,CTDNA}. Interest in these processes is due to applications in bio-medical fields and nano-science. In these investigations, fullerenes have been especially well studied, as they are an example of nano-particles with well defined structure and electronic properties \cite{campbell2000fullerene,C60GrapheneCT,nonadiabaticC60CT,ultrafastNanoTubeC60CT}.

In 1992 and 1993 Christian et al. \cite{Ne-C60CE,N-C60CE,O-C60CE,C-C60CE,Li/Na-C60} published a series of papers reporting relative cross sections for various fragmentation and charge transfer reactions involving $C_{60}$. These papers reported the relative values of charge transfer cross sections for collisions between carbon, nitrogen, neon, oxygen, lithium or sodium cations with $C_{60}$. These relative cross sections were measured as a function of collision energy for pure charge transfer:
\begin{equation}
    X^+ + C_{60} \rightarrow X + C_{60}^+ ,
\end{equation}
where X is either C, O, N, Ne, Na or Li. They also reported data on the energy dependence of relative cross sections for charge transfer and fragmentation collisions with $C_{60}$:
\begin{equation}\label{EqFragPath}
    X^+ + C_{60} \rightarrow X + C_{60-2n}^+ + n C_2 ,
\end{equation}
where $n$ is an integer. The cross section of $C_{59}$ formation was measured by Christian et al. to be significantly smaller then the cross sections for the fragmentation of an even number of carbon atoms (except in the case of the carbon cation). Authors suggested that the ions $X^+$ and $C_{60}$ fullerene form an intermediate long-living complex which decays into different reactive channels of fragmentation. For this scenario, each of the reactions shown in Eq.~\ref{EqFragPath} form a Breit-Wigner like resonance.

To our knowledge there has been no detailed theoretical exploration of these reactions for slow ion collisions. This is likely due to the difficulty surrounding theoretical approaches to dynamical properties of the $C_{60} + X^+$ system at different inter-particle distances. Even an accurate {\it ab initio} evaluation of the binding energy of $C_2$ in $C_{60}$ has already proven to be a difficult computational task \cite{boese1998c2,lifshitz2000c2}. In light of this, we have developed a semi-empirical model in order to offer theoretical insight into these charge transfer and fragmentation reactions while avoiding expansive {\it ab initio} calculations. Our model is essentially based on the semi-classical description of the $C_{60} + X^+$ system and on the models of non-adiabatic transitions developed in the theory of atomic collisions \cite{mott1949theory,child1996molecular,nikitin2012theory}.

We propose a simplified model which starts by using the jellium model suggested by Baltenkov et al. \cite{baltenkov2015jellium} to describe the electron wave functions in $C_{60}$. This model is combined with electrostatics and various measured physical parameters of $C_{60}$ to create pseudo-potential curves for a positive ion interacting with $C_{60}$. By using results of the resonant scattering theory of a multi-channel collision process and Landau-Zener theory of non-adiabatic transitions between different electronic states \cite{olson1970LZCT,olson1970LZStueck,zener1932OrigLZ}, we managed to create a theoretical model that fits to the original experimental data obtained for different projectile ions.

\section{Simplified Model of $C_{60}$ Electronic State}
Taking advantage of the symmetry of $C_{60}$, we use the jellium model suggested by Baltenkov et al. \cite{baltenkov2015jellium}. We consider charge transfer processes which involve only fullerene valence electrons.

Since $C_{60}$ is a sphere-like object we can approximate it as producing a potential well which is only a function of the radius $r$. At large distances $r \gg R_{C_{60}}$ ($R_{C_{60}}$ being the radius of the $C_{60}$ shell), the leading term of the real electron potential is the Coulomb potential. Therefore, one could write the potential as follows:
\begin{equation}
    U(r) = U^*(r) + U_{lr}(r) = U^*(r) - \frac{f(r,a)}{r}
\end{equation}
where $r$ is the distance of a valence electron from the center of $C_{60}$, $U^*(r)$ is the short range potential, $U_{lr}(r)$ is the long range potential and $f(r,a)$ is the Tang-Toennies damping function \cite{wuest2004potential,toennies1984DampingFunction}. The Tang-Toennies damping function prevents the long range Coulomb term from dominating at short ranges; it is given as:
\begin{equation}\label{EqTT}
f(r,a) = 1 - e^{-ar} \sum^{4}_{k=0} \frac{(a r)^k}{k!},
\end{equation}
where $a$ is an adjustable parameter that dictates the range at which the function is suppressed. For the short range part Baltenkov et al. asserts that a Lorentzian potential is more realistic than proposed alternatives (such as a delta function). This potential is given as:
\begin{equation}\label{EqJellPot}
    U^*(r) = -U_L\frac{d^2}{(r-R_{C_{60}})^2+d^2},
\end{equation}
where $U_L$ is the max well depth and the thickness of the well is given by $2d$. $R_{C_{60}}=6.665 ~a_0$\cite{C60Radius} corresponds to the radius of the $C_{60}$ cage. We consider only $s$-electrons with zero angular momentum. The $C_{60}$ wave function will be affected by the incoming ion and this affect can be different depending on the ion. To account for this the width of the positive background $d$ was used as a fit parameter for the density of states model, as discussed later in the article. The model pseudo-potential from Eq.~\ref{EqJellPot} provides adequate description of a single electron wave function in the region localized around the $C_{60}$ cage, if an accurate value of the electron energy $\epsilon$ is used as a parameter of the Schr\"{o}dinger equation. These wave functions will be used to calculate potentials for $C_{60}+X^+$, the long range behavior of these potential will be calculated using a more accurate method which does not use these wave functions (see Sect.~\ref{pot}). Therefore, we are most concerned with the behavior of the wave function around the $C_{60}$ shell and so we only use the potential shown in Eq.~\ref{EqJellPot} when calculating the wave functions. We solved the Schr\"{o}dinger equation numerically and obtained the electron wave function for an $\epsilon$ value equal to the experimental value of the photoionization energy for $C_{60}$, which is about $7.54 ~eV$ according to Hertel et al. \cite{hertel1992C60+energy}.

\begin{figure}
    \includegraphics[scale=0.4]{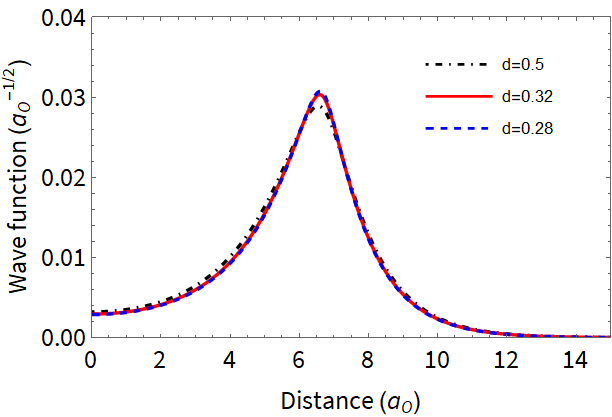}
    \caption{The electron wave functions calculated for $C_{60}$ using the jellium model. The black dotted and dashed curve is for $d=0.5~a_0$ and $U_L = 0.88 ~a_0$. The blue dashed curve is for $d=0.28 ~a_0$ and $U_L =1.18 ~a_0$. The red solid curve is for $d = 0.32 ~a_0$ and $U_L = 1.30 ~a_0$. The value of $d$ was chosen so the results of the density of states model (see Sect.~\ref{DofS}) matched data \cite{Ne-C60CE,N-C60CE,O-C60CE}.}
    \label{FigCWaves}
\end{figure}

We solved the Schr\"{o}dinger equation using a Verlet algorithm and adjusted the well depth $U_L$ until the wave function converges to zero as $r$ approaches infinity, indicating that $7.54 ~eVs$ had become an eigenvalue. We found that $d=0.5~a_0$, $d=0.32~a_0$ and $d=0.28~a_0$ obtained good results for the Oxygen ion, Nitrogen ion and Neon ion cross sections respectively (this will be discussed later). Since the energy is held constant, adjusting $d$ has a minimal effect on the wave functions. For $d=0.5~a_0$, $d=0.32~a_0$ and $d = 0.28~a_0$ we found $U_L = 0.88 ~a_0$, $U_L = 1.18 ~a_0$ and $U_L = 1.30 ~a_0$\footnote{The full $U_L$ values used for calculations are $U_L = 0.8798042104 ~a_0$, $U_L = 1.184260439 ~a_0$ and $U_L = 1.302635364 ~a_0$ for $d=0.5~a_0$, $d=0.32~a_0$ and $d = 0.28~a_0$ respectively} respectively. The resulting wave functions are shown in Fig.~\ref{FigCWaves}.

\section{Test of the electron wave functions: Resonant Charge Transfer}
\begin{figure}
    \includegraphics[scale=0.3]{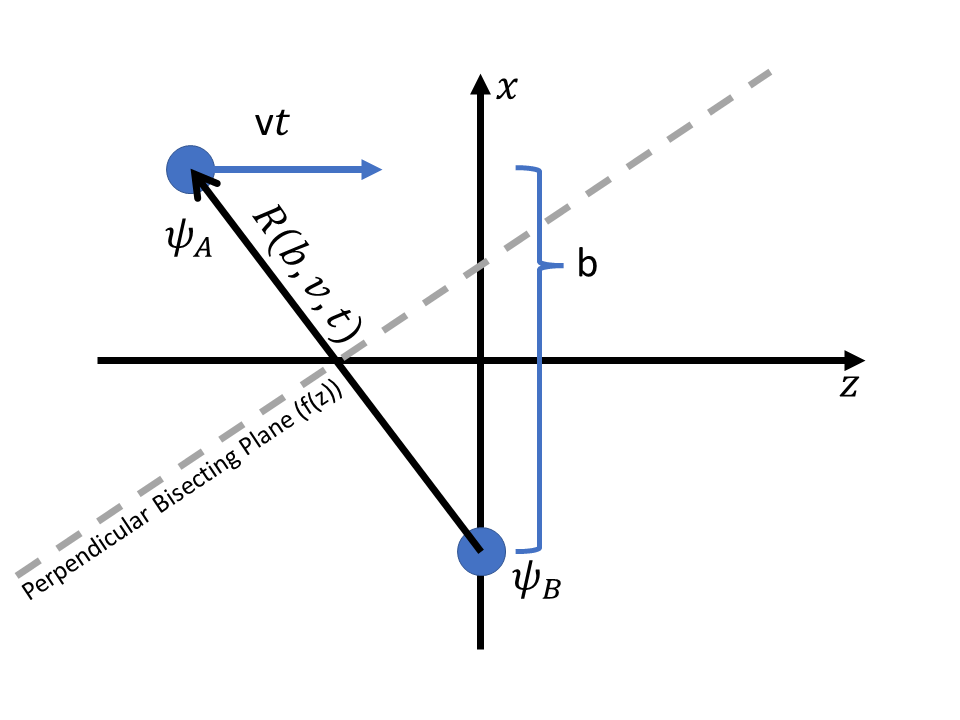}
    \caption{Above is a schematic of the method used to calculate the small angle charge transfer cross section for $C_{60}-C_{60}^+$. The unperturbed valence electron wave functions for the two $C_{60}$ molecules are represented by $\psi_A$ and $\psi_B$. $\psi_A$ and $\psi_B$ are the same wave functions except for a spacial shift. $\psi_A$ moves at a constant speed $v$ in the z-direction. The gray dashed line represents the plane halfway between the two $C_{60}$ molecules. This plane rotates as time passes.}
    \label{FigResCTSchem}
\end{figure}

We calculate the total cross sections for the resonant charge transfer case to test the accuracy of the one-electron wave functions obtained with the jellium model. The reaction for this case is: $C_{60}^+ + C_{60} \rightarrow C_{60} + C_{60}^+$. We use the Holstein-Herring method\cite{FluxPlaneTheory} as it simplifies the cross section so it only depends on the single particle electronic wave function. This method is briefly discussed below. A schematic of this method is shown in Fig.~\ref{FigResCTSchem}.

\begin{figure}
    \includegraphics[scale=0.4]{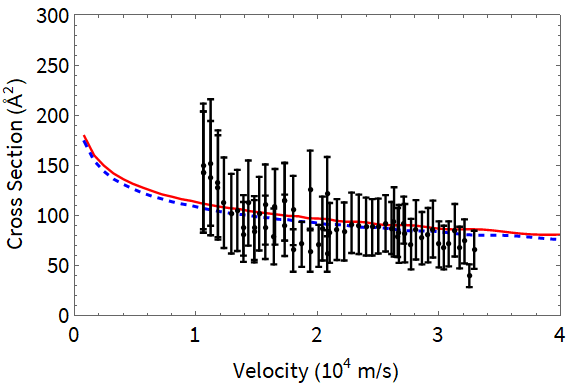}
    \caption{The small angle cross section for $C_{60}+C^+_{60}$ resonant charge transfer. The black dots represent data from Rohmund and Campbell \cite{rohmund1997C60ChargeTransfer}. The curves are theoretical calculations made using the wave functions from the jellium model, shown in Fig.~\ref{FigCWaves}. The Red line is for $d = 0.5 ~a_0$ and the blue dashed line is for $d = 0.28 ~a_0$. The adjustment of the parameter $d$ has little effect on the small angle cross section.}
    \label{FigResCT}
\end{figure}

The electrons of the $C_{60}-C_{60}^+$ pair move much faster then the colliding molecule. Therefore, the electron energies and wave functions adjust promptly to any changes of the inter-fullerene distance $R(b,v,t)$ (where $b$ is the impact parameter, $v$ is the velocity and $t$ is time). For the considered interval of collision energies the fullerene motion is classical, more specifically it is a straight line trajectory, if $R(b,v,t)$ is larger than twice the diameter of the $C_{60}$ carbon shell. The charge transfer cross section is calculated using the following:
\begin{equation}\label{EqBasic}
    \sigma = \int^\infty_0 2 \pi b~ P(b,v) ~db,
\end{equation}
where P(b,v) is the probability of charge transfer occurring.. Assuming a system of two $C_{60}-C_{60}^+$ states, which are degenerated as $R(b,v,t) \rightarrow \infty$, the $C_{60}-C_{60}^+$ valence electron wave function can be approximated as the gerade and ungerade sums of two $C_{60}$ valence electron wave functions $\psi({\bf r})$. If we also assume an adiabatic process, the probability of the resonant charge transfer collisions $P(v,b)$ can be calculated using the energy splitting between the gerade and ungerade quasi-molecular states:
\begin{equation}
    P(v,b)= \sin^2{\int_{-\infty}^{\infty}\frac{\epsilon_g-\epsilon_u}{2}dt},
\end{equation}
in atomic units. This energy splitting can be estimated using the electron probability flux through the plane half way between the two $C_{60}$ molecules i.e:
\begin{eqnarray}
\epsilon_g - \epsilon_u =2 \int^{\infty}_{-\infty}\int^{\infty}_{-\infty}~\psi(f(z),y,z)\nonumber\\
\times
\nabla \psi(f(z),y,z)\cdot \frac{\vec{R}(b,v,t)}{||\vec{R}(b,v,t)||}dy~dz,\label{Splitting}
\end{eqnarray}
where $\psi({\bf r})$ is the valence electron wave function of an unperturbed $C_{60}$ molecule. The above equation is again in atomic units. For lowest order perturbation theory, we can use the jellium wave functions. More accurate theoretical models take into account a wave function transformation induced, at large distances, by the ion. We have chosen the impact parameter $b$ to be along the x-axis and the two particles to be shifted by $\pm b/2$ from the origin. The moving projectile was chosen to move along the z-axis at a constant speed $v$ starting at $z=-\infty$. The $f(z)$-function characterizes the plane that is the perpendicular bisector for the two $C_{60}$ molecules (The gray dashed line in Fig. \ref{FigResCTSchem}). In this case, $f(z)$ is given as:
\begin{equation}\label{EqFz}
    f(z) = -\frac{v t}{b} z + \frac{v^2 t^2}{2 b}.
\end{equation}
This plane rotates with time and at $t=0$ it is equivalent to the y-z plane. We would like to study only elastic collisions with charge transfer so the impact parameter has a minimum of $2R_{C_{60}}$. If $b<2R_{C_{60}}$ then the two $C_{60}$ cages would collide resulting in other collision processes and non-straight line trajectories. The resulting calculated cross section compared to data reported by Rohmund and Campbell for small angle scattering \cite{rohmund1997C60ChargeTransfer} is shown in Fig.~\ref{FigResCT}. We repeated this calculation using two different values for the parameter $d$. The changing of $d$ made a small difference in the cross section, this indicates that the parameter $d$ can be adjusted. The calculations are in good agreement with the measured small angle cross sections indicating that the jellium model wave functions are adequate. It is important to note that Glotov and Campbell later corrected the total cross sections reported by Rohmund and Campbell \cite{rohmund1997C60ChargeTransfer} by taking in account large scattering angles \cite{glotov2000CorRes}. The theory we used assumes straight line trajectories so the original data on small angle charge transfer cross sections is more applicable. The presence of large angle scattering in resonant charge transfer collisions has never been observed in ion-atom collisions or in collisions involving simple molecules. A new theoretical model should be developed to describe resonant charge transfer collisions with large scattering angles.

\section{Pseudo-potentials}\label{pot}
\begin{figure}
    \includegraphics[scale=0.3]{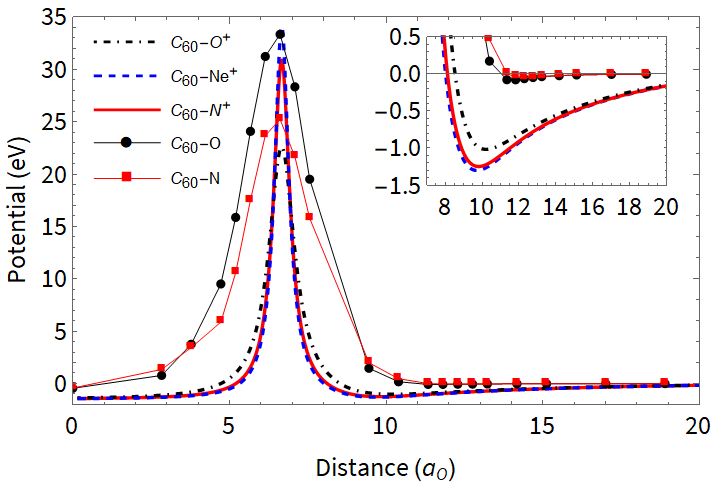}
    \caption{The potential constructed for $C_{60}$ interacting with a positive ion. The curves are our model potentials and the points represent DFT calculations. The wave function used for the Oxygen, Nitrogen and Neon potentials are shown in Fig.~\ref{FigCWaves}. The long range behavior of the potentials are shown in the inset.}
    \label{FigCPotentails}
\end{figure}

A simplified description of collision processes involving complex system can be achieved with the introduction of pseudo-potentials\cite{pseudopotentialElectronic,pseudopotentialSolidState}. A very large number of quasi-molecular states are involved in the considered charge transfer and fragmentation processes. We have constructed pseudo-potentials which describe resonance scattering and charge transfer process as an evolution of $C_{60}+X^+$ quasi-molecular states. The long-range asymptotic behavior of our potentials corresponds to the long range interactions in $C_{60}+X^+$ and $C_{60}^{+}+X$ systems. Modeling molecular interactions using electrostatic potentials is a well known technique, which is especially effective for long-range interactions. We combine this technique with the wave functions calculated using the jellium model to calculate the potential energies for $C_{60}$ interacting with a positive ion. To this end, we convert the wave functions shown in Fig.~\ref{FigCWaves} to charge distributions and add it to the interaction energy of the incoming cation with the jellium positive background. For the long range behavior, $C_{60}$ is treated as a dielectric sphere. The charged ion induces an image point and a line charge inside $C_{60}$ \cite{lindell1992eDielecSph}. This image is responsible for a weak attractive force outside the spherical shell. The dielectric constant was set to $\epsilon_r = 4$ based on results by Ren et al \cite{ren1991Dielectric} and Ortiz-Lopez et al \cite{ortiz2003dielectric}. At large distances, this attractive polarization roughly behaves as a $r^{-4}$ potential. To prevent the long range potential from dominating at shorter ranges the long range potential is multiplied by a Tang-Toennies damping function \cite{wuest2004potential} which is shown in Eq.~\ref{EqTT}. In this case $a$ is chosen so the long range potential has as little effect on the short range potential as possible while keeping the long range well as deep as possible. The potential created by the dielectric approximation is much stronger than the typical approach which approximates the long range well as being a result of the polarizability of the two molecules. The resulting potential is depicted in Fig.~\ref{FigCPotentails} with the long range behavior shown in the inset.

For comparison with the pseudo-potentials, density functional theory (DFT) calculations were performed using the recently developed APFD dispersion corrected hybrid functional\cite{APFD} and the 6-31+G(d) gaussian basis set. This method was designed to provide an accurate description of long range molecular interactions. A comparison of the pseudo-potentials and DFT potentials for $C_{60}$ interacting with Oxygen and Nitrogen is shown in Fig.~\ref{FigCPotentails}. $C_{60}+O^+$ is not the ground state of the system and therefore it is difficult to perform a DFT calculation of this configuration, which is why we performed the potential energy curve calculation for $C_{60}+O$. Both the DFT calculations and the psuedo-potentials predict a high peak at $6.665 ~a_0$ and two shallow wells: one within the $C_{60}$ shell, and one directly outside the $C_{60}$ shell. All DFT calculations were performed with the Gaussian 16 electronic structure programs\cite{g16}.

Although not directly included in the calculation of the $C_{60}$-ion interaction, the electron wave function is slightly disturbed by the presence of the ion. Such influence can be accurately determined at large distances $r$ between the projectile ion and $C_{60}$, but requires elaborate numerical computation. The most simple empirical way to take into account the charge's interaction with the electronic state's of $C_{60}$ is to include a dependence of the width parameter $d$ in Eq.~\ref{EqJellPot} on the projectile.

\begin{figure}
    \includegraphics[scale=0.35]{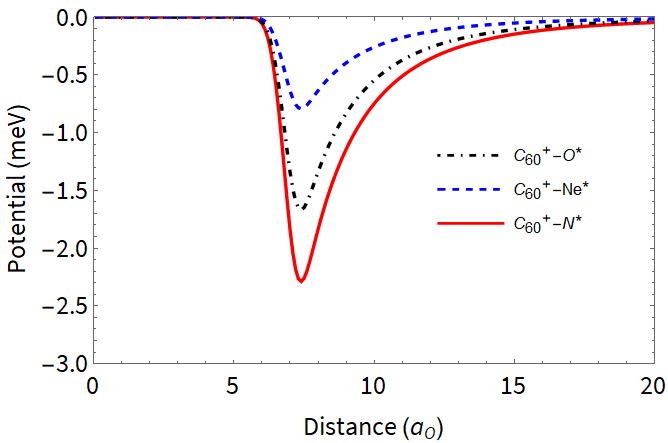}
    \caption{The potentials constructed for $C_{60}^+$ interacting with Oxygen (black dashed and dotted line), Nitrogen (red line) and Neon (blue dashed line). The potential was calculated using the polarizabilities of the atoms interacting with a positively charged spherical shell. To avoid non-differentialable points, the potentials where made smooth by applying a Gaussian Filter. This potential should be accurate for longer ranges.}
    \label{FigCpPotentails}
\end{figure}

Quasi-molecular charge transfer theory also requires potentials for $C_{60}^+$ interacting with a neutral atom in both ground and excited states in order to locate regions where adiabatic charge transfer can occur. At large distances between the $C_{60}^+$ molecule and the atom, the charge around the $C_{60}^+$ molecule induces a dipole in the neutral atom. The magnitude of this dipole is dependent on the polarizability of the neutral atom. The long-range potential $V(r)$ of the interaction between $C_{60}^+$ and the neutral atom with polarizability $\alpha_X$ behaves as: $V(r)= - \alpha_{X}/ 2 r^4$. A set of potentials for the $C_{60}^+ + X$ interaction (with X representing O, N, and Ne) has been computed. These potentials are shown in Fig.~\ref{FigCpPotentails}. The polarizabilities $\alpha$ used are: $\alpha = 0.802 ~\text{\normalfont\AA}^3$ for Oxygen \cite{miller1978Polar,nesbet1977ExcitedPolarizabilities}, $\alpha = 1.10 ~\text{\normalfont\AA}^3$ for Nitrogen \cite{miller1978Polar,nesbet1977ExcitedPolarizabilities} and $\alpha = 0.382 ~\text{\normalfont\AA}^3$ for Neon \cite{olney1997PolarNeon}. The polarizabilities vary depending on the state but do not vary significantly, so for simplicity we only used the ground state polarizabilities. These potentials are likely not accurate inside or near the spherical shell but should be more accurate on longer ranges since the polarization potential is the dominating term for large distances. Since the relevant regions of non-adaibatic interaction (which are needed for Landau-Zener charge transfer theory) between the two potentials are located at long ranges, these potentials should be accurate enough in the region of interest.

\section{Non-Resonant Charge Transfer and fragmentation}
\begin{figure*}
    \includegraphics[scale=0.34]{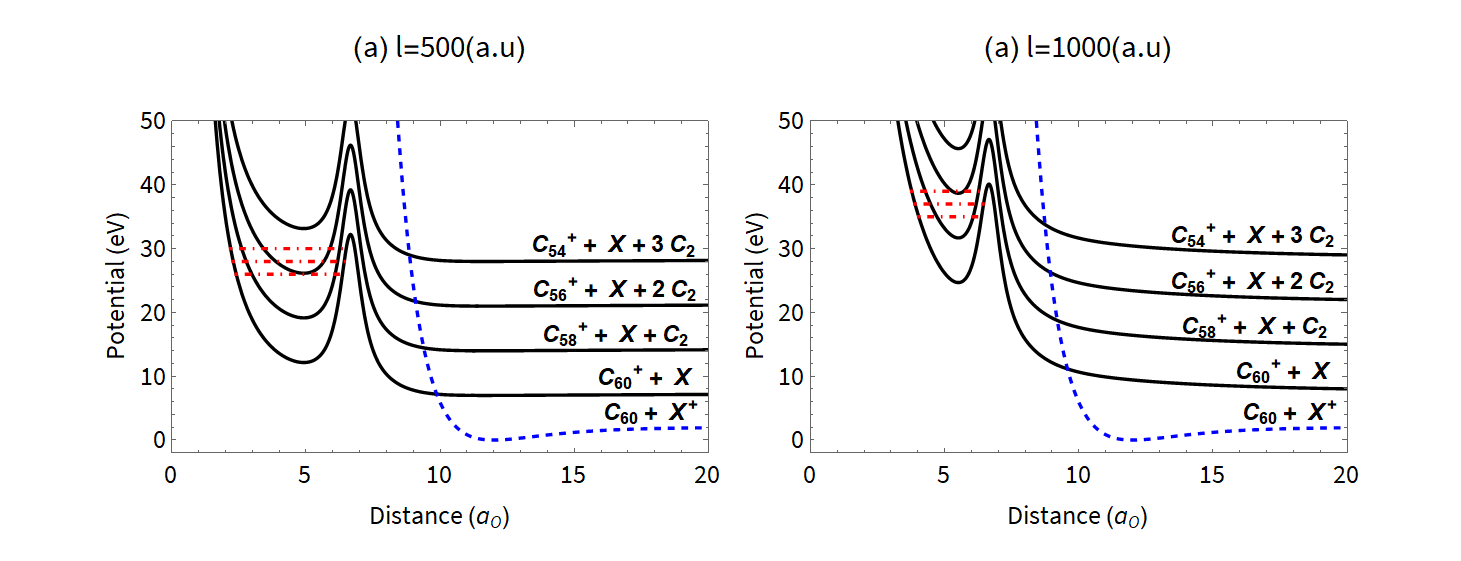}
    \caption{Above are two sketches representing our model. The black lines represent potentials for $C_{60-2n}^++X+nC_2$, where $n$ is 0,1,2 or 3. These potentials represent the lower energetic boundary for the reaction since $nC_2$ can have any amount of kinetic energy. The blue dashed line represents the potential for the initial state $C_{60}+X^+$. The gap between these levels is between $5-7 ~eVs$, to be inline with estimates for the energy of the $C^+_{60} \rightarrow C^+_{58} + C_2$ reaction \cite{C60massSpect}. Landau-Zener theory assumes that the transferring between states only occurs in the region where these two potentials cross. The red dashed and dotted line represents internal bound states for $C_{60}^++X$. The energy of these bound states change for different values of the angular momentum $l$ (corresponding to different impact parameters). When the collisional energy is roughly equal to one of these bounds states the chance of tunneling increases causing a resonance.}
    \label{FigSchematic}
\end{figure*}

Christian et al.\cite{Ne-C60CE,N-C60CE,O-C60CE} proposed that the fragmentation occurs separately from charge transfer. They proposed that when charge transfer occurs the resulting $C_{60}^+$ is in a rovibrational state that allows the process to be near resonant. Since $C_{60}$ has many rovibarational states there is very likely a state which allows for this. This rovibrational state is proposed to be unstable and the $C_{60}^+$ then decays into $C_{60-2n}^++nC_2$. We used the potentials constructed in the previous section to create a reaction pathway model that can explain uniquely the experimental set of data on the charge transfer collisions between slow ions and $C_{60}$ \cite{Ne-C60CE,N-C60CE,O-C60CE,C-C60CE}. We treat the charge transfer and the fragmentation of $C_{60}$ separately (as Christian et al. proposed), multiplying the results from both together to obtain the final results. The fragmentation is treated as an inelastic scattering process using a density of states model. Charge transfer is modeled using Landau-Zener charge transfer theory. A schematic representation of our model is shown in Fig.~\ref{FigSchematic}. The curves in Fig.~\ref{FigSchematic} represent effective potentials of different reactive channels. The asymptotic difference between these channels is about $5-7~eVs$, to be inline with estimates for the energy of the $C^+_{60} \rightarrow C^+_{58} + C_2$ reaction \cite{C60massSpect}. According to Landau-Zener theory, it is near the intersection between these channels that the state may change.

\subsection{Density of States}\label{DofS}
We estimate parameters of the states involved in the resonance scattering process. The potential shown in Fig.~\ref{FigCPotentails} is part of the effective potential of radial motion. The full potential of the collision contains the centrifugal potential which depends on the impact parameter $b$. This term affects the depth and shape of the interior potential well. For large $b$, the total potential becomes completely dominated by the centrifugal term; this occurs somewhere around $b \approx 8~a_0$ depending on the projectile and its kinetic energy. The interior potential well was fit to a harmonic oscillator for several values of $b$ and the top 20 states were calculated each time \cite{connor1968analytical,connor1968semi}. A list of states and their energies were obtained for all values of $b$ that still allowed for an interior well. When a particle collides with roughly the same energy as one of these states a resonance occurs. Each resonance is broadened using a Gaussian function and all the states are summed together to obtain the multiple resonance shape function $\sigma^*(\epsilon)$, shown below:
\begin{equation}
    \sigma^*(\epsilon) = \sum_i \frac{1}{c\sqrt{2\pi}}\exp{[-\frac{1}{2}(\frac{\epsilon^*_{i}-\epsilon}{c})^2]},
\end{equation}
where $c$ is the standard deviation of the Gaussian (which corresponds to the effective width of each state), the sum is over all the bound state energies (represented by $\epsilon^*_i$) calculated previously. The shape function $\sigma^*(\epsilon)$ is normalized to the total number of bound states $\epsilon^*_{i}$. Broadening the energies using a Gaussian function has been used to take into account a variety of physical effects such as: deformation and vibrational excitation of the $C_{60}$ cage, Doppler broadening, lifetime broadening and other possible broadening processes. The standard deviation $c$ is the same for all states and is fit to the data from Christian et al. The entire function is scaled so the peak of the model is the same height as the peak of the data. This last step must be done since Christian only measured the relative cross sections.

\begin{figure}
    \includegraphics[scale=0.35]{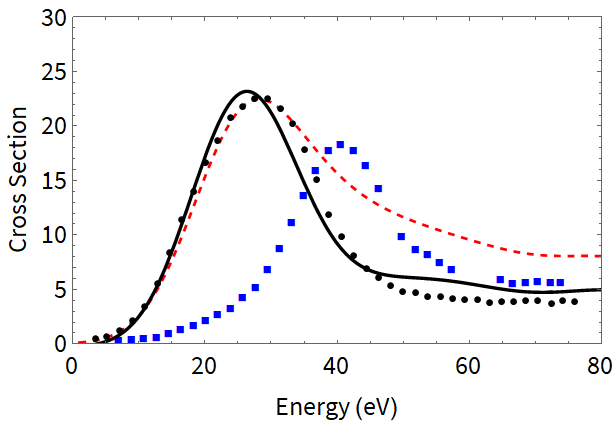}
    \caption{The plot constructed using our density of states model as well as the measured cross sections for $C_{60}+O^+$ \cite{O-C60CE}. The black dots represent data for an oxygen ion colliding with $C_{60}$ and knocking off one $C_2$ leaving $C_{58}$. The blue squares represent data for an oxygen ion colliding with $C_{60}$ and knocking off two $C_2$ leaving $C_{56}$. The red dashed line represents our model before accounting for a decrease in cross section due to the opening of the $C_{56}+2C_2$ reaction channel. The black line represents the model after accounting for this reduction in cross section.}
    \label{FigDenOfStates}
\end{figure}

The resulting shape functions for $C_{60}$ interacting with an oxygen cation are shown in Fig.~\ref{FigDenOfStates} along side the data digitized from Christian's paper \cite{O-C60CE}. In this example the width of the Gaussian $c$ was found to be $7.39 ~eVs$.

Notice the peak of our potential shown in Fig.~\ref{FigCPotentails} has roughly the same energy as the peak of the fragmentation of a single $C_2$. Due to the size of $C_{60}$ a wide range of impact parameters will only shift the barrier up slightly. Most of the states calculated using our method have an energy slightly higher than this peak, causing a high density of states at the energy of the peak. Since each state corresponds to a resonance, a high density of energy levels creates a large peak in the cross section. Therefore, a peak in the potential barrier translates to a peak in the cross section. The curve over predicts the cross section for a specific fragmentation for higher energies. This is consistent with Christians explanation that the probability of one interaction is reduced through the opening of new reaction pathways causing a reduction in the cross section. To account for this we subtracted one cross section from the next and refit our cross sections for our final figures.

\subsection{Landau-Zener Model of Interaction between Different Collisional Channels}\label{SectLZ}
Landau-Zener model of non-adiabatic transitions between different states of compound systems assumes that the transfer between states is only possible when the energy of the two states are roughly the same \cite{olson1970LZCT,olson1970LZStueck,zener1932OrigLZ}. This model was chosen because of its effectiveness and simplicity. When the two potentials get close to each other the interaction between different diabatic states creates an avoided crossing. It is only at these avoiding crossings that the diabatic state can change. Every time the projectile reaches the distance of a given crossing there is some probability of transitioning between states. Given $N$ number of states each with only one crossing with the original state, (Fig.~\ref{FigSchematic} shows this scenario with $N = 4$) the probability of finishing in a particular state is:
\begin{eqnarray}
    P_n = p_n \prod_{i=1}^{n}(1-p_i)[1+(\sum^{N}_{j=n+1}p_j^2\prod_{k=n+1}^{j-1}(1-p_k)^2)\nonumber\\+\prod_{m=n+1}^{N}(1-p_m)^2],\label{EqGen}
\end{eqnarray}
where $P_n$ is the probability of finishing in state $n$ and $p_n$ is the probability of remaining in the initial diabatic states at the crossing $n$. The probability to remain in the initial diabatic state at each one of the individual crossing points has been shown to be \cite{olson1970LZCT,olson1970LZStueck,zener1932OrigLZ}:
\begin{equation}
    p =e^{- 2 v_x/v_l},\label{EqProbCross}
\end{equation}
where $v_x$ is the characteristic velocity and $v_l$ is the velocity at the turning point. The characteristic velocity is defined as:
\begin{equation}
    v_x=\frac{\pi V^2_{12}}{|V^{'}_{11}(R_x)-V^{'}_{22}(R_x)|},\label{EqCharVel}
\end{equation}
$V^{'}_{11}(R_x)$ and $V^{'}_{22}(R_x)$ are the derivatives of the two potentials at the location of the crossing ($R_x$ being the position of the crossing point). $V_{12}$ is the interaction term of the Hamiltonian. $V_{12}$ is difficult to calculate and is used as a fit parameter.

When the electron has been captured by the cation it doesn't necessarily occupy any particular electronic state. Therefore, we calculate crossing points for multiple excited states. This leads to a system with many adiabatic regions predicted by the crossing points of the two potentials. The position of the regions of adiabatic transitions have been predicted using the potentials calculated previously, shown in Fig.~\ref{FigCpPotentails} and Fig.~\ref{FigCPotentails}. To simplify the calculations required for our model, the potentials in Fig.~\ref{FigCpPotentails} are used multiple times and are shifted up so that the energy at infinite distance is equal to the excited state energy. These shifted potentials represent different electronic excited states. The excited state energies for Oxygen, Nitrogen and Neon are all taken from \cite{moore1993Ospectrum}.

\begin{figure}
    \includegraphics[scale=0.3]{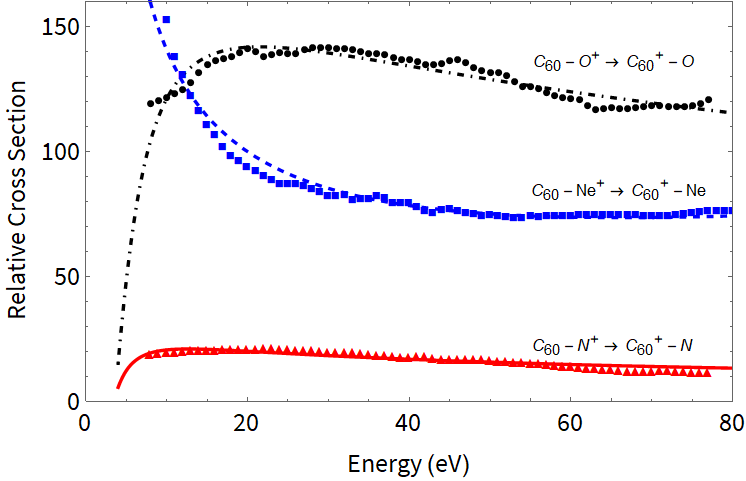}
    \caption{The curves represent Landau-Zener charge transfer theory for two crossings where both interaction terms of the Hamiltonian were fit parameters. The crossing points were predicted from the potentials shown in Fig.~\ref{FigCPotentails} and Fig.~\ref{FigCpPotentails}. The points represent data taken by Christian et al. \cite{Ne-C60CE,N-C60CE,O-C60CE}. The cross sections for all the different fragmentation reactions were summed together to produce the points shown.}
    \label{FigCTAll}
\end{figure}

Some of the crossings are located within the spherical shell. In order to access these inner crossings the ion would have to penetrate $C_{60}$ without destroying it. The $C_{60}$ structure is not maintained during the interaction (since at least one $C_2$ is removed) and so the inner crossings were ignored. We also found that only two states are sufficient to get a reasonable fit indicating that the resulting neutral atom is most likely in one of two states. Landau-Zener charge transfer theory predicts that the first crossing dominates the reaction and the impact of each subsequent crossing gets lower and lower. Therefore, the excited state with the lowest energy that still has more energy than the photoassociation energy of $C_{60}$ ($7.54 ~eVs$) is overwhelmingly the most likely state the electron transfers into. For two crossings Eq.~\ref{EqGen} reduces to:
\begin{eqnarray}
 P_1 = p_1 (1-p_1)(1+p_2^2+(1-p_2)^2),
\\
 P_2 = 2 p_2 (1-p_2) (1-p_1).
\end{eqnarray}
Because the data does not differentiate between different electronic states of the resulting outgoing neutral atom, the total charge transfer cross section is the cross section of the two states added together. The interaction term of the Hamiltonian ($V_{12}$) of the two crossings and an overall magnitude parameter are all fit to the data. Since we are assuming that charge transfer and fragmentation are separate interactions, all the reactions start with pure charge transfer. Therefore, the total cross section for just charge transfer ($C^+_{60}+X\rightarrow C_{60}+X^+$) is the sum of the cross sections for all the charge transfer and fragmentation reactions. The pure charge transfer cross section calculated for Oxygen, Nitrogen and Neon are all shown in Fig.~\ref{FigCTAll}.

Notice that the fundamental form of the Neon charge transfer cross section is different from the other two ions. This decaying exponential like form fits well to crossing points with negative energy while the Nitrogen and Oxygen curves fit well to positive energy crossings. Our method predicts negative energy crossings only so they were shifted up by the photoionization energy of $C_{60}$ to create positive energy crossings for Nitrogen and Oxygen.

\section{Results}
\begin{figure}
    \includegraphics[scale=0.3]{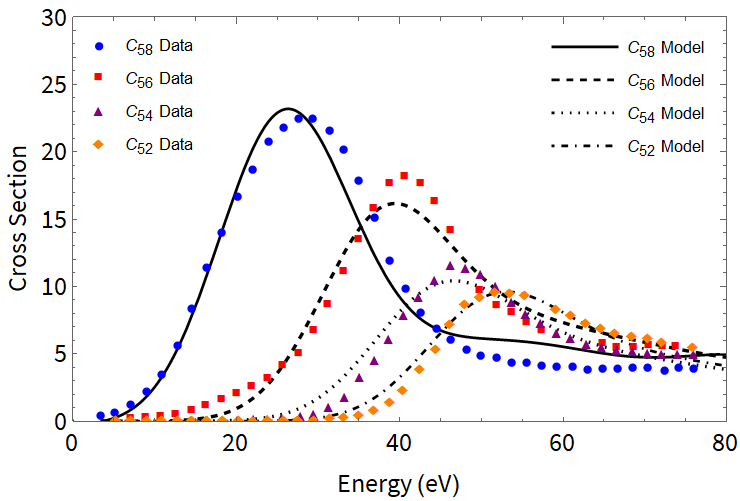}
    \caption{The data from Christian et al. \cite{O-C60CE} for $C_{60}$ colliding with $O^+$ plotted along side our model (shown in black).}
    \label{FigFinalO}
\end{figure}

\begin{figure}
    \includegraphics[scale=0.3]{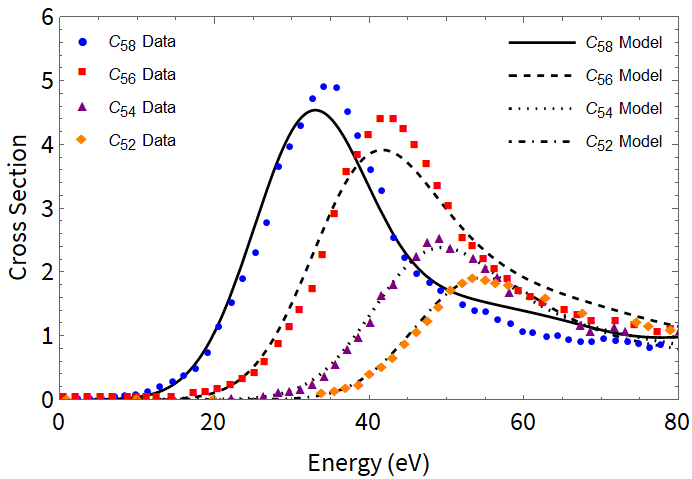}
    \caption{The data from Christian et al. \cite{N-C60CE} for $C_{60}$ colliding with $N^+$ plotted along side our model (shown in black).}
    \label{FigFinalN}
\end{figure}

\begin{figure}
    \includegraphics[scale=0.3]{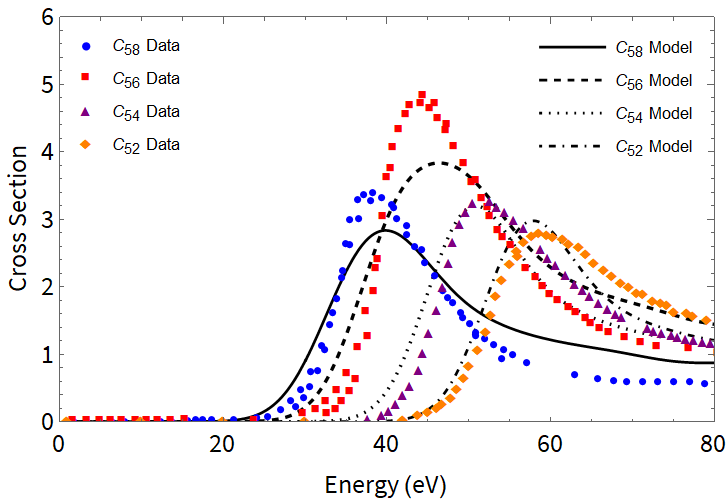}
    \caption{The data from Christian et al. \cite{Ne-C60CE} for $C_{60}$ colliding with $Ne^+$ plotted along side our model (shown in black).}
    \label{FigFinalNe}
\end{figure}

To obtain cross sections for $C_{56}$, $C_{54}$ and $C_{52}$, the potentials were shifted up in energy so that the peak matched the location of the peak for the other reactions. The charge transfer cross sections shown in Fig.~\ref{FigCTAll} are multiplied by the density of states shape function ($\sigma^*(\epsilon)$). As previously mentioned in Sect.~\ref{SectLZ}, the opening of new channels takes away probability from previous channels. To account for this mechanism, the final step was to fit the data for $C_{58}$ to:
\begin{equation}
    \sigma_{C_{58}f}=a_1 ~ \sigma_{C_{58}} - b_1~ \sigma_{C_{56}},
\end{equation}
$\sigma_{C_{58}}$ and $\sigma_{C_{56}}$ are the cross sections from the previous step and $a_1$ and $b_1$ are fit parameters. $a_1$ and $b_1$ are related to the probabilities ($P_n$ in Eq.~\ref{EqGen}) of finishing in the single $C_2$ fragmentation or double $C_2$ fragmentation states. This was repeated for the other cross sections. The final results for Oxygen, Nitrogen and Neon are shown in Fig.~\ref{FigFinalO}, Fig.~\ref{FigFinalN} and Fig.~\ref{FigFinalNe} respectively.

Our model is consistent with statements made by Christian et al. \cite{O-C60CE}. Charge transfer occurs first leaving the $C_{60}^++O$ complex in an unstable excited state. This unstable state then decays expelling some number of $C_2$ molecules. The cross section of all the fragmentation reactions are roughly constant and the cross section gets reduced with the opening of subsequent channels.

This model makes a few predictions about these fragmentation and charge transfer processes. Electrostatic forces create a high barrier at the radius of $C_{60}$, this barrier must be overcome for fragmentation to become likely. If the barrier is overcome with enough energy, more carbon atoms will be knocked loose. This barrier remains mostly unchanged for various impact parameters, which results in the resonances shown in our model. This implies the physical size of $C_{60}$ plays a key role in these reactions. Landau-Zener charge transfer implies that the electron will most likely end up in the lowest excited state of the atom that still has more energy then the photoionization energy of $C_{60}$. Each subsequent energy level will have a smaller and smaller probability. Finally, the form of the pure charge transfer cross section is explained by Landau-Zener charge transfer theory. A positive energy crossing point creates a resonant like cross section, like what is shown for Oxygen and Nitrogen. A negative energy crossing point creates an exponential decay like cross section, like the one shown for Neon.

\section*{Acknowledgments}
The work of MB and JS on this project was supported via the UCLA grant ONRBAA13-022. RC was supported by the National Science Foundation (NSF) Grant No. PHY-2034284. HRS and VK acknowledge support from the NSF through a grant for ITAMP at Center for Astrophysics $|$ Harvard \& Smithsonian. One of the authors (DV) is also grateful for the support received from the National Science Foundation through grants PHY-1831977 and HRD-1829184.

\bibliography{references}

\end{document}